\documentclass[
floatfix,
aps,
prb,
twocolumn,
superscriptaddress,
]{revtex4}

\usepackage{amssymb}
\usepackage{latexsym}
\usepackage[dvips]{graphicx}
\usepackage{amsmath}
\usepackage[mathcal]{eucal}
\usepackage{graphicx}
\usepackage{dcolumn}
\usepackage{amsfonts}
\usepackage{bm}
\usepackage{epsfig}
\usepackage{times}

\newcommand{\be}{\begin{equation}}
\newcommand{\ee}{\end{equation}}
\newcommand{\bea}{\begin{eqnarray}}
\newcommand{\eea}{\end{eqnarray}}

\def\nn{\nonumber}
\def\dingle{\lambda}

\def\eph{\epsilon}

\def\pph{\varphi}

\def\oc{\omega_{\mbox{\scriptsize {c}}}}

\def\vs{v_s}

\def\os{\omega_{s}}
\def\olo{\omega_{\mbox{\scriptsize {o}}}}

\def\tq{\tau_{\mbox{\scriptsize {q}}}}
\def\ttr{\tau}

\newcommand{\req}[1]{Eq.\,(\ref{#1})}

\newcommand{\rfig}[1]{Fig.\,\ref{#1}}

\newcommand{\rref}[1]{Ref.\,\onlinecite{#1}}

\begin{document}
\title{
Phase of phonon-induced resistance oscillations in a high-mobility two-dimensional electron gas
}

\author{A.\,T. Hatke}
\affiliation{School of Physics and Astronomy, University of Minnesota, Minneapolis, Minnesota 55455, USA}

\author{M.\,A. Zudov}
\email[Corresponding author: ]{zudov@physics.umn.edu}
\affiliation{School of Physics and Astronomy, University of Minnesota, Minneapolis, Minnesota 55455, USA}

\author{L.\,N. Pfeiffer}
\affiliation{Princeton University, Department of Electrical Engineering, Princeton, NJ 08544, USA}

\author{K.\,W. West}
\affiliation{Princeton University, Department of Electrical Engineering, Princeton, NJ 08544, USA}


\begin{abstract}
We report on experimental studies of magnetoresistance oscillations that originate from the resonant interaction of two-dimensional electrons with thermal transverse-acoustic phonons in very high-mobility GaAs/AlGaAs quantum wells.
We find that the oscillation maxima consistently occur when a frequency of a phonon with twice the Fermi momentum {\em exceeds} an integer multiple of the cyclotron frequency.
This observation is in contrast to to all previous experiments associating resistance maxima with magnetophonon resonance and its harmonics.
Our experimentally obtained resonant condition is in excellent quantitative agreement with recent theoretical proposals.
\end{abstract}
\maketitle

Resonant interaction of a two-dimensional electron gas (2DEG) with thermal \emph{optical} phonons was predicted\citep{gurevich:1961} and observed\citep{tsui:1980} in magnetotransport measurements in GaAs/AlGaAs heterostutures a long time ago.
Experimentally, this interaction gives rise to a rather weak enhancement of the longitudinal resistivity occurring whenever the phonon frequency $\olo$ equals an integer multiple of the cyclotron frequency, $\oc=eB/m^\star$, where $B$ is the magnetic field and $m^\star$ is the effective electron mass.
In GaAs,\citep{waugh:1963} $\olo \sim 10^{13}$ s$^{-1}$ and observation of the resulting magnetoresistance oscillations requires both high temperatures $T\gtrsim 10^2$ K and strong magnetic fields $B \gtrsim 10^2$ kG.

About a decade ago, another class of magnetoresistance oscillations was discovered\citep{zudov:2001b} in a high-mobility ($\mu_e\sim 10^{6}$ cm$^{2}$/Vs) 2DEG at much lower temperatures $T \sim 1-10$ K and magnetic fields $B \sim 1-10$ kG.
These so-called phonon-induced resistance oscillations (PIROs) are currently understood in terms of resonant interaction of 2D electrons with thermal \emph{acoustic} phonons.
We note that such interaction has also been observed in experiments on phonon drag long time ago.\citep{zavaritskii:1983}

Resonant electron-acoustic phonon scattering in a 2DEG is made possible by virtue of a special selection rule, which results in electron {\em backscattering} off of an acoustic phonon.\citep{zudov:2001b}
In this scenario, even though phonons of many different energies are present, {\em only} the most energetic phonons that electrons can scatter off can be considered to describe the effect.
Such phonons carry momentum $\simeq 2k_F$ ($k_F$ is the Fermi wavenumber), and, as a result, their frequency $\os$ is determined entirely by the parameters of the 2DEG, i.e. the sound velocity $\vs$ and the electron density $n_e$,
\be
\os = 2 k_F \vs\,.
\label{os}
\ee
Due to absorption and emission of $2k_F$ phonons, an electron undertakes \emph{indirect} transitions between Landau levels, and the longitudinal resistivity acquires a periodic in $1/B$ oscillatory correction.
Similar to magnetoresistance oscillations originating from optical phonons,\citep{tsui:1980} this correction oscillates with the ratio of the $2k_F$ phonon frequency $\os$ to the cyclotron frequency, 
\be
\eph = \frac{\os}{\oc}\,,
\label{eph}
\ee
and can be described by 
\be
\delta \rho \simeq A(\eph) \cos (2\pi\eph + \pph),~~\eph=\frac \os \oc\,.
\label{eq.piro.1}
\ee

In contrast to other low-field transport phenomena, such as microwave-induced\citep{ryzhii:1970,ryzhii:1986,zudov:2001a,ye:2001,durst:2003,lei:2003,dmitriev:2003,willett:2004,mani:2004e,dmitriev:2005,dorozhkin:2005,studenikin:2005,lei:2005a,yang:2006,lei:2006c,zhang:2007c,dmitriev:2007b,studenikin:2007,hatke:2008a,hatke:2008b,khodas:2008,dmitriev:2009b,lei:2009,hatke:2009a,fedorych:2010,khodas:2010,dai:2010,hatke:2011b,hatke:2011c} and Hall field-induced\citep{yang:2002,bykov:2005c,zhang:2007a,lei:2007,vavilov:2007,hatke:2009c,hatke:2011a} resistance oscillations (known to evolve into zero-resistance\citep{mani:2002,zudov:2003,yang:2003,andreev:2003,zudov:2004,smet:2005,auerbach:2005,zudov:2006a,zudov:2006b,bykov:2006,wiedmann:2010b,andreev:2011,dorozhkin:2011} and zero-differential resistance states,\citep{bykov:2007,hatke:2010a,wiedmann:2011a},repectively), PIRO have not received much attention until recently,\citep{zhang:2008,zudov:2009,hatke:2009b,bykov:2009a,bykov:2010a,lei:2008,raichev:2009,dmitriev:2010b,raichev:2010b} and several unsolved issues remain.

One such issue is the value of the phase $\pph$ appearing in \req{eq.piro.1}.
All of the existing experiments\citep{zudov:2001b,yang:2002b,bykov:2005b,zhang:2008,zudov:2009,hatke:2009b,bykov:2009a,bykov:2010a} associated PIRO maxima and minima with integer and half-integer $\eph$ values, respectively, suggesting that $\pph \approx 0$.
On the other hand, recent theories\cite{lei:2008,raichev:2009,dmitriev:2010b} predict that the phase is not necessarily zero and is determined by the specifics of the phonon spectrum, the width of the quantum well, and the temperature, which controls excitation of different phonon modes.
However, to date, no experimental study exists that examines these theoretical predictions.

In this Rapid Communication, we systematically investigate the period and the phase of phonon-induced resistance oscillations in several very high-mobility ($\mu_e > 10^{7}$ cm$^{2}$/Vs) 2DEGs. 
An extremely low level of disorder in our samples allows reliable measurement of up to eight oscillations at relatively low temperatures.
The latter condition is rather crucial to ensure that contribution from more energetic longitudinal phonon modes is sufficiently weak and can be neglected.
Analysis of the oscillation waveform reveals that, in all of our samples, the dominant mode has a velocity $\vs \approx 3.4$ km/s, suggesting a dominant contribution from the transverse branch of the phonon spectrum. 
More importantly, in contrast to previous experimental studies which assumed $\pph = 0$, we find that all of the observed oscillations are best described by a nonzero phase $\pph \approx - 0.24\,\pi$ in all of the samples studied.
We discuss our findings in terms of recent theoretical proposals\citep{raichev:2009,dmitriev:2010b} and find excellent quantitative agreement.

All of our samples (A, B, and C) are lithographically defined Hall bars fabricated from very high-mobility symmetrically doped GaAs/AlGaAs quantum well structures grown by molecular-beam epitaxy.
The electron density $n_e$ (in units of $10^{11}$ cm$^{-2}$), the disorder-limited mobility $\mu_e$ (in units of $10^7$ cm$^2$/Vs), and the quantum well width (in nanometers), were 3.7, 1.0, and 30\,(sample A), 3.0, 1.2, and 30\,(sample B), and 2.9, 2.4, and 28\,(sample C), respectively.
Magnetoresistivity $\rho(B)$ was measured in a $^{3}$He cryostat using low-frequency lock-in technique.

\begin{figure}[t]
\includegraphics{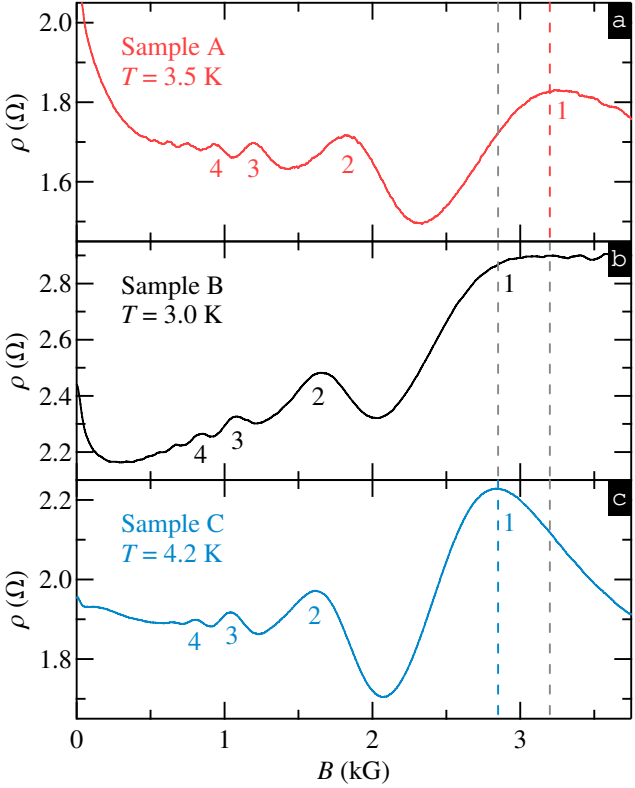}
\caption{(Color online) Magnetoresistivity $\rho(B)$ measured in (a) sample A at $T=3.5$ K, in (b) sample B at $T=3.0$ K, and in (c) sample C at $T=4.2$ K. 
Integers mark the oscillation order $n$.
Vertical lines are drawn to illustrate higher magnetic field of the first PIRO peak in sample A (right line) than in sample B (left line).
}
\label{fig1}
\end{figure}
In \rfig{fig1}, we present the magnetoresistivity $\rho(B)$ measured in (a) sample A at $T=3.5$ K, in (b) sample B at $T=3.0$ K, and in (c) sample C at $T=4.2$ K.
All samples reveal several orders (up to $n=8$ in sample A) of pronounced PIROs, which are marked by integers.
Oscillations become progressively stronger with the magnetic field reflecting the $A\propto\lambda^2$ scaling of the oscillation amplitude. 
Here, $\lambda = \exp(-\pi/\oc\tq)$ is the Dingle factor, and $\tq$ is the quantum lifetime.\citep{hatke:2009b,raichev:2009,dmitriev:2010b}
 
Close examination of \rfig{fig1} reveals that PIROs occur at somewhat higher magnetic fields in sample A (cf.\, right vertical line) compared to samples B and C (cf.\,left vertical line).
This difference is well accounted for by $\os\propto k_F \propto \sqrt{n_e}$ scaling of the phonon frequency, see \req{os}.
Finally, \rfig{fig1} demonstrates that PIROs appear regardless of the sign and the magnitude of the magnetoresistance effect. 
Indeed, oscillations are superimposed on a slowly varying background, which is decreasing, increasing, and roughly constant in samples A, B, and C, respectively.

\begin{figure}[t]
\includegraphics{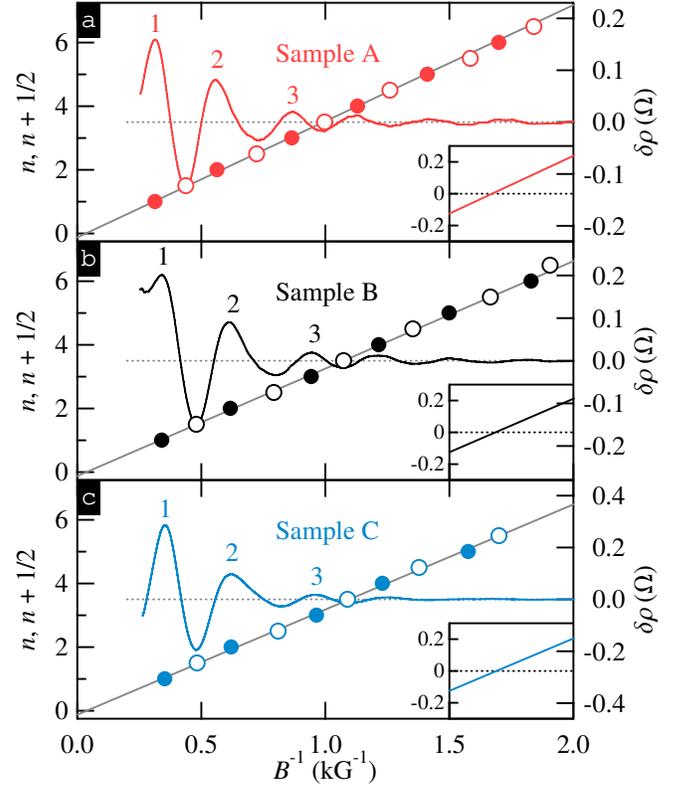}
\caption{(Color online)
Oscillation order $n$ (solid circles) for the maxima  and $n+1/2$ (open circles)  for the minima (left axis) and the oscillatory part of the resistivity $\delta\rho$ (right axis) vs $1/B$ measured in (a) sample A, (b) sample B, and (c) sample C.
Lines are linear fits to the oscillation order $j$.
Insets illustrate that the fits extrapolate to nonzero value at $1/B=0$ for all samples. 
}
\label{fig2}
\end{figure}
To examine the period of the oscillations, we extract the oscillatory part of the magnetoresistivity $\delta \rho$ by subtracting the background from the data presented in \rfig{fig1}.
The result of such extraction is presented in \rfig{fig2} (right axis) clearly showing periodicity in $1/B$ for all three samples in agreement with \req{eq.piro.1}. 
Following earlier experimental studies,\citep{zudov:2001b,yang:2002b,zhang:2008,zudov:2009,hatke:2009b,bykov:2009a,bykov:2010a} we associate the maxima and the minima of $\delta\rho$ with integer $n$ and half-integer $n+1/2$ values, respectively.
These values for the maxima (solid circles) and the minima (open circles) are plotted in \rfig{fig2} (left axis) as a function of $1/B$.
For all samples studied, we observe universal linear dependence on $1/B$ for both the maxima and the minima.
From the slope of the linear fits (solid lines), we obtain the value of the sound velocity $\vs \approx 3.4$ km/s for all three samples, again confirming $\os \propto \sqrt{n_e}$ scaling.
This value is close to the velocity of the bulk transverse acoustic mode $v_{\rm T}=3.35$ km/s.

We further notice that none of the fits pass through zero at $1/B=0$.
As shown in the insets of \rfig{fig2}, all of the fits reveal a {\em negative} intercept at $-\delta \approx - 0.12$, suggesting that the positions of the PIRO maxima$^{(+)}$ and minima$^{(-)}$ in all of our samples can be described by
\be
\eph^+ = n + \delta ,\,\,\eph^- = n + \frac 1 2 + \delta ,\,\,\delta \approx 0.12\,.
\label{eq.piro.max.1}
\ee
From this value of the intercept $\delta$ we obtain $\pph \approx - 0.24\,\pi$ appearing in \req{eq.piro.1}. 
\begin{figure}[t]
\includegraphics{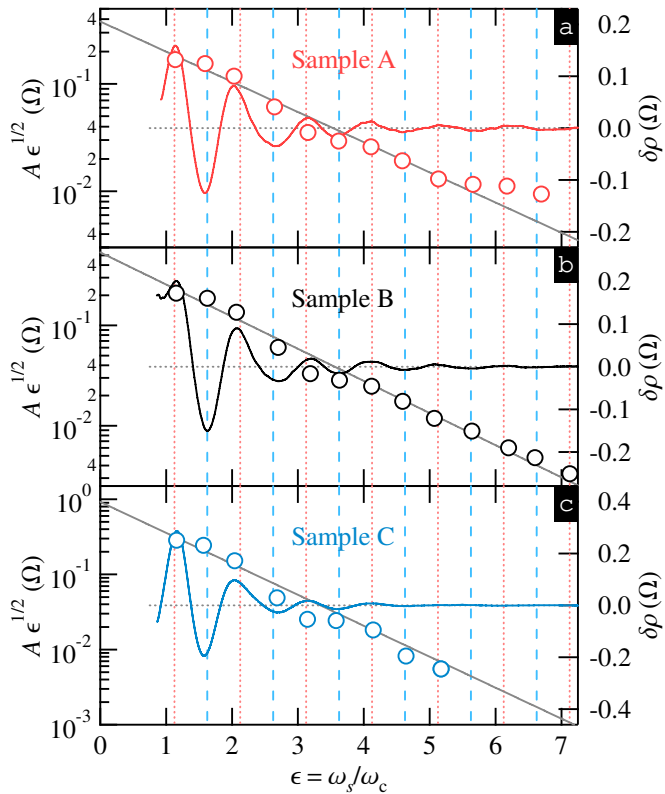}
\caption{(Color online)
Normalized PIRO amplitude $A\eph^{1/2}$ (left axis) and the oscillatory part of the resistivity $\delta\rho$ (right axis) vs $\eph=\os/\oc$ measured in (a) sample A, (b) sample B, and (c) sample C.
Solid lines are fits to the data.
Dotted (dashed) vertical lines are drawn at $\eph^+ = n+1/8$ ($\eph^- = n+5/8$).
}
\label{fig3}
\end{figure}

To verify the validity of \req{eq.piro.max.1} for individual maxima and minima, we convert $1/B$ to $\eph$ and present the oscillatory part of the resistivity $\delta\rho$ as a function of $\eph=\os/\oc$ in \rfig{fig3} (right axis).
Plotted in this way, the data confirm that the period is equal to unity and that {\em all} maxima (minima), in all of our samples, are {\em consistently} shifted from integer (half-integers) $\eph$ to higher values.
Except for a few deviations, the value of the shift is roughly the same for all extrema as demonstrated by vertical dotted (dashed) lines drawn at $\eph^+=n+1/8$ ($\eph^-=n+5/8$) for the maxima (minima).
This observation further confirms that PIROs in all of our samples, under the conditions of the experiment, can be well described considering a single-phonon mode.

We now discuss our experimental findings in terms of recent theoretical proposals.\cite{raichev:2009,dmitriev:2010b}
The first paper,\cite{raichev:2009} considering {\em anisotropic} phonon dispersion, identified important contributions from three phonon modes $-$ piezoelectric transverse mode polarized along $[001]$ (out-of-plane) direction and two deformational longitudinal modes polarized along $[100]$ and $[110]$ (in-plane) directions.
Analytical expressions for partial contributions from these modes, obtained for the case of narrow quantum wells $k_F b \lesssim 1$ and in the limit of $\eph \gg 1$, revealed the phase $\pph$ of $-\pi/2$, $\pi/4$, and $-\pi/4$, respectively.\cite{n1}
However, in all of our samples, $k_F b \gtrsim 4$ and numerical calculations,\cite{raichev:2009} performed for sample A used in the present Rapid Communication, revealed a considerably {\em reduced} phase of the transverse contribution for {\em all} of the experimentally detected oscillation orders [see, Fig.\,3\,(b) in \rref{raichev:2009}].
While in this calculation of \rref{raichev:2009} the phase does approach $\pph = - \pi/2$, it happens only at very high oscillation orders $\eph^{1/2} \sim k_F b$, not accessible in our experiment.

In a more recent paper,\cite{dmitriev:2010b} it was argued that for the case of wide quantum wells $k_F b \gg 1$, only phonons with very small out-of-plane momentum $q_\perp \sim b^{-1} \ll k_F$ interact with electrons, provided that $\eph$ is not too large $\eph^{1/2} \lesssim k_Fb$. \cite{n2}
As a result, the electron-phonon-scattering form factor can be approximated by a $\delta$ function, and in the limit of high temperature $k_BT \gg \hbar\oc,\,\hbar\os$,\cite{n3} the oscillatory part of the resistivity is given by
\bea
\frac {\delta \rho}{\rho} &=&  \frac {g^2k_B T \ttr} \hbar \dingle^2 [J_0(2\pi\eph)-J_2(2\pi\eph)]\nn\\
&\simeq& \frac {2g^2 k_B T \ttr} {\pi\hbar\sqrt{\eph}} \dingle^2 \cos (2\pi\eph - \pi/4).
\label{eq.piro}
\eea
Here, $g$ is the dimensionless electron-phonon coupling-constant,\cite{n4} and $\ttr$ is the transport-scattering time.
The second line of \req{eq.piro} is justified for $2\pi\eph \gg 1$, i.e., it works well already at $\eph = 1$.  
Direct comparison with \req{eq.piro.1} clearly reveals $\pph = - \pi/4$, which is in excellent quantitative agreement with our experimental result $\pph \simeq - 0.24\,\pi$.

Finally, we analyze the oscillation amplitude $A$ on $\eph$ in terms of \req{eq.piro}, which predicts 
$A\propto\eph^{-1/2}\lambda^2 = \eph^{-1/2}e^{-\alpha\eph}$, where $\alpha = 2\pi/\tq\os$.
In \rfig{fig3} (left axis), we present reduced oscillation amplitude $A\,\eph^{1/2}$ as a function of $\eph$ and observe roughly exponential dependence, which further confirms the dominant contribution of a single-phonon mode in our experiments.
The deviations from this dependence are likely due to subleading contributions from other modes, e.g., interface phonons, originally proposed to explain PIROs\citep{zudov:2001b,ponomarev:2001} and/or weakly excited longitudinal modes,\citep{raichev:2009,dmitriev:2010b} which have subleading contributions under the conditions of our experiment.

To summarize, we have investigated the period and the phase of phonon-induced resistance oscillations in a very high-mobility 2DEG.
Due to a very low level of disorder in our samples, we were able to reliably measure up to eight oscillations at relatively low temperatures, which ensured a dominant contribution of the transverse-acoustic phonon mode.
The extracted value of the sound velocity is in good agreement with the known value of $v_{\rm T} = 3.35$ km/s.
Furthermore, in all of the samples studied, we found that the oscillations exhibit a nonzero phase $\pph \approx - 0.24\,\pi$.
This value is in excellent agreement both with the analytical expression,\citep{dmitriev:2010b} obtained for the experimentally relevant case of wide quantum well and not too large $\eph$, and with numerical calculations.\cite{raichev:2009} 

We thank I. Dmitriev and O. Raichev for discussions.
The work at Minnesota was supported by NSF Grant No. DMR-0548014.
The work at Princeton was partially funded by the Gordon and Betty Moore Foundation as well as the NSF MRSEC Program through the Princeton Center for Complex Materials (DMR-0819860).


\begin{thebibliography}{73}
\bibitem[{n1()}]{n1}
\bibinfo{note}{Longitudinal modes become relevant at higher temperatures than those used in our experiment.}
\bibitem[{n2()}]{n2}
\bibinfo{note}{For $\eph^{1/2} \gg \max\{1, k_F b\}$, the phase of PIROs should approach $\pph = -\pi/2$, regardless of the quantum well width.}
\bibitem[{n3()}]{n3}
\bibinfo{note}{While this condition is not satisfied in our experiment, its violation is not expected to affect the phase.}
\end{thebibliography}
\end{document}